\documentstyle[sprocl]{article}

\input epsfig.sty
\begin{document}
\title{DIJET CROSS SECTIONS AND RAPIDITY GAPS BETWEEN JETS IN PHOTOPRODUCTION AT HERA}

\author{M. HAYES}
\author{On behalf of the ZEUS Collaboration}

\address{H.~H. Wills Physics Laboratory, Tyndall Avenue, Bristol BS8 1TL
UK}

\maketitle\abstracts{
Dijet cross sections from the original 1994 ZEUS data analysis are shown compared to NLO calculations. New 1995 cross sections are shown compared to Monte Carlo predictions. The fraction of dijet events with a rapidity gap inbetween the jets is also presented for the new 1995 data.
}

%\section{Introduction}

In $e^+p$ scattering at HERA, an almost real photon ($P^2 \approx 0$, where $P^2$ is the negative four momentum squared of the photon) can be emitted from the positron with momentum fraction $y$ and collide with the proton. At leading order (LO) there are two distinct types of process: (a) where all of the photon interacts directly with the proton (direct process) and (b) where the photon resolves into partons which interact with the proton (resolved process). Both processes can produce two high $E_T$ jets which provide a hard scale, meaning that they should be calculable in perturbative QCD. In this contribution we report on two different dijet cross section measurements.
Another possibility is resolved hard scattering with the exchange of a colour singlet object. A measurement of the fraction of rapidity gap events compared to normal dijet events is presented, which probes these processes.

The present results are based on 2.6 $\rm{pb}^{-1}$ of the data taken during 1994 HERA running and on 5.7 $\rm{pb}^{-1}$ from the 1995 data. In both years HERA was running with 820 GeV protons and 27.5 GeV positrons.

Photoproduction processes are separated from other processes by the requirement that no positron is found in the main detector. This corresponds to a cut of $P^2 < 4 \rm{GeV}^2$ and gives a median $P^2\approx 10^{-3} \rm{GeV}^2$.

Jets are located using the energy deposit pattern in the uranium-scintillator calorimeter. Results presented here use both the cone type and the clustering type of jet finding algorithms. Two types of cone algorithm (EUCELL and PUCELL) are used in ZEUS. Both involve moving a cone around in $\eta-\phi$ \ \footnote{ $\eta = - \ln\tan (\theta/2)$, where $\theta$ is polar angle measured from the proton direction and $\phi$ is the azimuthal angle.} space. Both comply with the snowmass convention. The clustering algorithm (KTCLUS \cite{KTCLUS}) uses an $E_T$ recombination scheme to build jets from `clusters' of nearby particles.

\section{Dijet Cross Sections}

\begin{figure}
\begin{center}
\epsfig{file=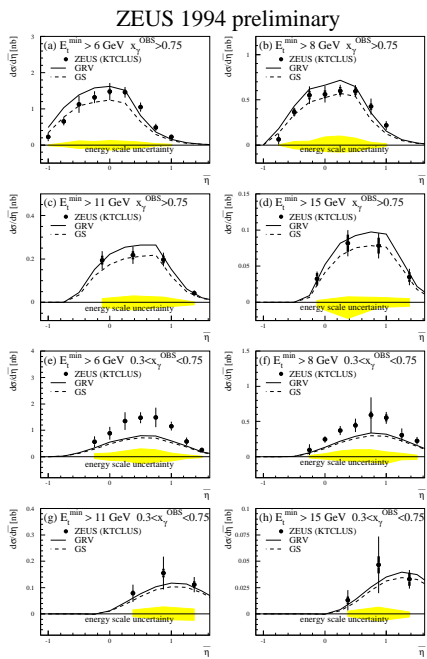,width=6cm}
\caption{Dijet cross sections $d\sigma/ d\overline\eta$. The thin (thick) error bars represent systematic (statistical) errors. The shaded band is the uncertainty in the calorimeter energy scale.}
\label{dijet94}
\end{center}
\end{figure}

The first set of cross sections presented are based on the 1994 HERA running. Dijet events are found as detailed above using the clustering algorithm. The jets are also required to lie in the range $-1.375 < \eta^{jet} < 1.875$. The cut $\Delta\eta = |\eta_1-\eta_2| < 0.5$ is applied to increase the correlation with the parton momenta. The cross section was measured as a function of $\overline\eta = (\eta_1+\eta_2)/2$, which is the boost of the dijet system in the lab..

The $e^+p$ cross sections $d\sigma/ d\overline\eta$ are presented in figure \ref{dijet94}integrated over $E_T^{jet} > 6,8,11,15$ GeV in two regions of $x_\gamma^{OBS}$\ 
\footnote{
In LO the fraction of photon's momentum taking part in the scattering can be estimated using
$x_\gamma^{OBS} = {\sum_{jets} E_T^{jet}e^{-\eta^{jet}} / (2 y E_e)}$
where $E_e$ is the energy of the incoming positron.
}
: (a)-(d) $x_\gamma^{OBS} > 0.75$ corresponding to mostly direct and (e)-(h) $0.3 < x_\gamma^{OBS} < 0.75$ corresponding to mostly resolved. The cross sections are compared to next-to-leading-order (NLO) calculations by Klasen and Kramer \cite{Klasen}. The calculations are shown for two photon structure function sets, GRV and GS.
For $x_\gamma^{OBS} > 0.75$ the NLO calculations agree well with the data for $E_T > 8$ GeV although at present differentiation between photon structure functions is not possible. For the lowest $E_T$ graph, $\overline\eta < 0$ agrees with the GS set, but at higher $\overline\eta$ the data agrees with the GRV set. For $0.3 < x_\gamma^{OBS} < 0.75$ the data lies above the calculation for the first two graphs. However the agreement is good for $E_T > 11$ GeV. Again no differentiation between photon structure functions is possible.

\begin{figure}
\begin{center}
\epsfig{file=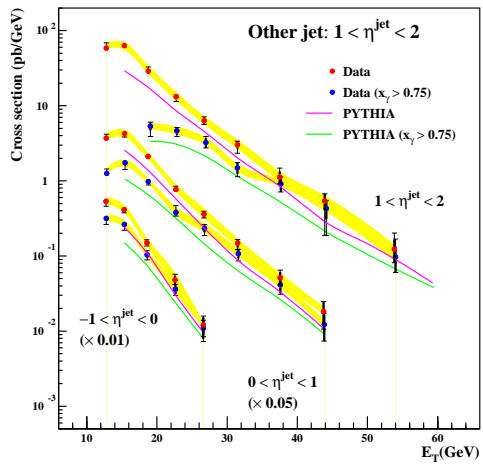,width=7.0cm}
\caption{ZEUS Preliminary 1995 dijet cross sections $d\sigma /d E_T$; one of the jets is required to be forward ($1<\eta^{jet}<2$).
In the 1995 plots the thick error bars are statistical and the thin ones are statistical and systematic added in quadrature. The shaded band represents the uncertainty in the calorimeter energy scale.}
\label{dijet95:1}
\end{center}
\end{figure}

The second set of cross sections presented are based on the 1995 HERA running. Dijet events are found as above using a cone algorithm. The jets are also required to lie in the range $-1 < \eta^{jet} < 2$. Both jets are required to have $E_T > 11$ GeV. The following graphs are all symmetrized in $\eta_{jet}$, both jets being plotted twice for each of their $\eta$ values. $E_T$ always refers to the highest $E_T^{jet}$\ \cite{Aurenche}.

The $e^+p$ cross sections $d\sigma /d E_T$ are presented in figure \ref{dijet95:1}. One of the jets is required to be forward ($1<\eta^{jet}<2$), and the three sets of points show entries for the other jet being in three separate parts of the detector ($1<\eta^{jet}<2$, $0<\eta^{jet}<1$ or $-1<\eta^{jet}<0$). A subset of the data ($x_\gamma^{OBS} > 0.75$) is also plotted separately. PYTHIA curves are also plotted for each set of points.
The graph shows that at high $E_T$ the dominant contribution is from direct processes. The Monte Carlo shape appears to agree well with the data, although the overall normalization is low.

%The same data points are shown in the group of four graphs on the right in figure \ref{dijet95:1}. However the points are scaled by a factor of $(E_T)^N$ where $N$ is chosen to enhance the shape differences between the monte carlo and data. Only the entire data set is now plotted and is shown against both PYTHIA and HERWIG. 

\begin{figure}
\epsfig{file=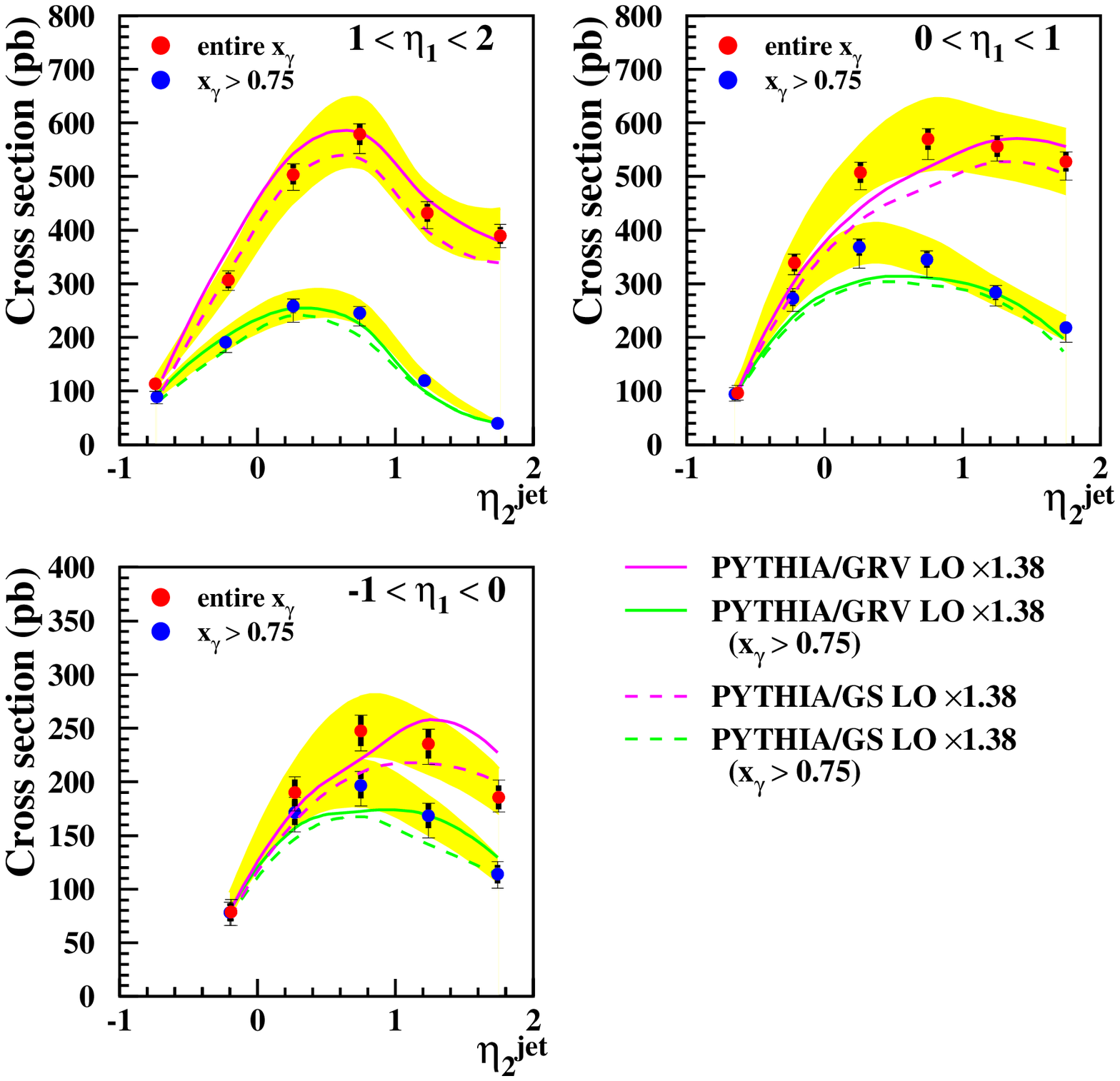,width=5.9cm}
\epsfig{file=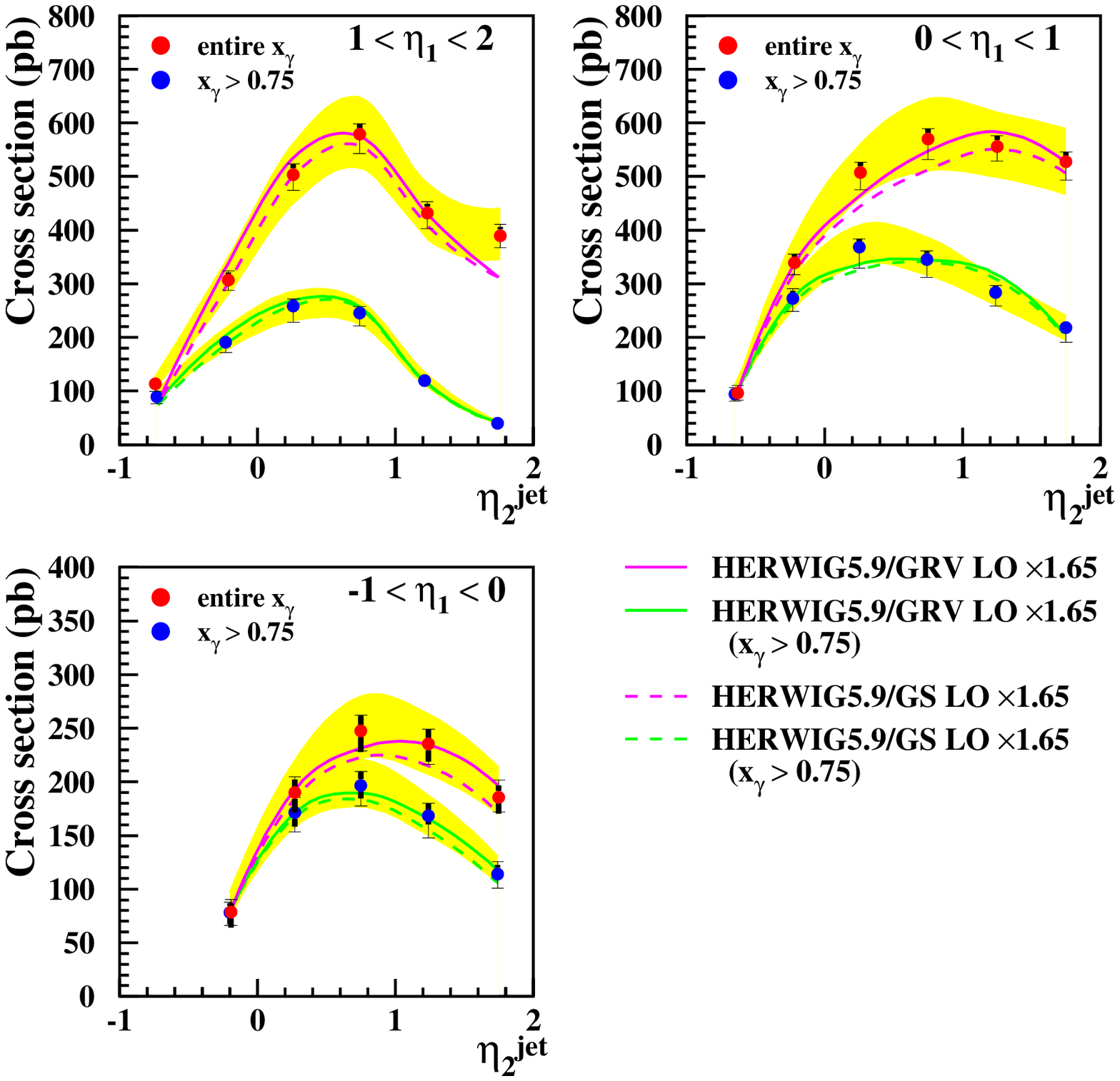,width=5.9cm}
\caption{ZEUS preliminary 1995 dijet cross sections $d\sigma /\eta_2$. The extra cut $E_T > 14$ GeV has been applied to the leading jet. Errors as in figure \ref{dijet95:1}.}
\label{dijet95:3}
\end{figure}

The $e^+p$ cross sections $d\sigma /\eta_2$ are presented in figure \ref{dijet95:3}. The graphs show the pseudorapidity of the other jet, given that one jet is in a definite area of the detector ($1<\eta^{jet}<2$, $0<\eta^{jet}<1$ or $-1<\eta^{jet}<0$). The extra cut $E_T > 14$ GeV has been applied to the leading jet. Again, the entire data set is plotted as well as the subset $x_\gamma^{OBS} > 0.75$. These are compared to two PYTHIA curves [left plots] corresponding to the two photon structure function sets, GRV and GS. The PYTHIA lines have been scaled, by eye, to the data. HERWIG is shown in the right set of plots.
Both Monte Carlos show good agreement with the data when considering the calorimeter uncertainty scale. However, PYTHIA appears more forward when requiring one jet to be in the middle or rear. HERWIG agrees well with the shape formed by the data points.

\section{Rapidity Gaps Between Jets}

It was shown\cite{gaps:photon95} that there exists a class of dijet events with large separation between jets with no significant energy deposits between the jet cone edges. Standard direct and resolved processes can produce these `gap' events by random particle multiplicity fluctuations. However, these are expected to become less likely exponentially with increasing gap width ($\Delta\eta$). A surplus of events above this exponential decrease is described by the exchange of an unknown colour singlet.

\begin{figure}
\begin{center}
\epsfig{file=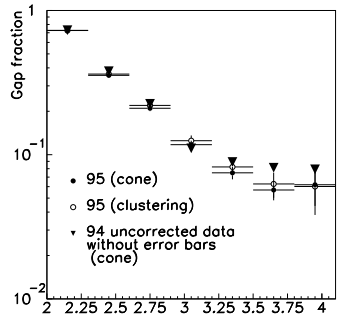,width=4.8cm}
\caption{Uncorrected preliminary data showing the gap fraction. Error bars represent statistical errors. The gap fraction based on 1994 data is also shown for comparison.}
\label{gaps}
\end{center}
\end{figure}

Figure \ref{gaps} shows the fraction of dijet events with a gap (the gap fraction), i.e. the number with a gap divided by the total number of dijet events. An $E_T^{jet} > 6$ GeV was used with both types of jet finders. A cut of $\overline\eta < 0.75$ is also applied. A gap is defined as the absence of a cluster of calorimeter cells with total transverse energy of greater than 250 MeV. 
The gap fractions agree well with both jet finders and the 1994 data published result. A levelling out of the gap fraction is seen for $\Delta\eta > 3.0$, rather than an exponential fall off, indicating the exchange of a colour singlet.

\end{document}